\documentclass[final,3p,times]{elsarticle}
\usepackage{lipsum}
\makeatletter
\def\ps@pprintTitle{%
 \let\@oddhead\@empty
 \let\@evenhead\@empty
 \def\@oddfoot{}%
 \let\@evenfoot\@oddfoot}
 
\usepackage{amssymb}
\usepackage{amsmath}
\usepackage{xurl}
\usepackage{hyperref}
\usepackage{subcaption}

\journal{Computer Methods in Applied Mechanics and Engineering}

\begin{document}

\begin{frontmatter}

\title{High-throughput digital twin framework for predicting neurite deterioration using MetaFormer attention}

% Use letters for affiliations, numbers to show equal authorship (if applicable) and to indicate the corresponding author
\author[a]{Kuanren Qian}
\author[b]{Genesis Omana Suarez}
\author[b]{Toshihiko Nambara}
\author[b]{Takahisa Kanekiyo}
\author[a,c]{Yongjie Jessica Zhang}

\address[a]{Department of Mechanical Engineering, Carnegie Mellon University, 5000 Forbes Ave, Pittsburgh, PA 15213, USA}
\address[b]{Department of Neuroscience, Mayo Clinic, 4500 San Pablo Road, Jacksonville, FL 32224, USA}
\address[c]{Department of Biomedical Engineering, Carnegie Mellon University, 5000 Forbes Ave, Pittsburgh, PA 15213, USA}

\begin{abstract}
Neurodevelopmental disorders (NDDs) cover a variety of conditions, including autism spectrum disorder, attention-deficit/hyperactivity disorder, and epilepsy, which impair the central and peripheral nervous systems. 
Their high comorbidity and complex etiologies present significant challenges for accurate diagnosis and effective treatments.
Conventional clinical and experimental studies are time-intensive, burdening research progress considerably.
This paper introduces a high-throughput digital twin framework for modeling neurite deteriorations associated with NDDs, integrating synthetic data generation, experimental images, and machine learning (ML) models. 
The synthetic data generator utilizes an isogeometric analysis (IGA)-based phase field model to capture diverse neurite deterioration patterns such as neurite retraction, atrophy, and fragmentation while mitigating the limitations of scarce experimental data. 
The ML model utilizes MetaFormer-based gated spatiotemporal attention architecture with deep temporal layers and provides fast predictions.
The framework effectively captures long-range temporal dependencies and intricate morphological transformations with average errors of 1.9641\% and 6.0339\% for synthetic and experimental neurite deterioration, respectively.
Seamlessly integrating simulations, experiments, and ML, the digital twin framework can guide researchers to make informed experimental decisions by predicting potential experimental outcomes, significantly reducing costs and saving valuable time. 
It can also advance our understanding of neurite deterioration and provide a scalable solution for exploring complex neurological mechanisms, contributing to the development of targeted treatments.
\end{abstract}


\end{frontmatter}

\section{Introduction}

Neurodevelopmental disorders (NDDs) encompass a diverse collection of conditions, from intellectual disabilities and attention-deficit/hyperactivity disorder (ADHD) to severe, debilitating illnesses such as schizophrenia and bipolar disorder~\cite{thapar2017neurodevelopmental}. 
In the United States, recent studies estimate that by eight years of age, 25\% of children with public insurance and 11\% with private insurance are diagnosed with NDDs~\cite{straub2022neurodevelopmental}. 
These conditions are further complicated by high rates of comorbidity, presenting significant challenges to accurate diagnosis and effective treatment planning~\cite{morris2020neurodevelopmental}.
Advances in neurodevelopmental research provide new opportunities to unravel the complexities of the developing brain and identify key factors influencing its growth~\cite{d2017neurodevelopmental, kawatani2024human}. 
However, the combinatorial complexity of potential factors and limited understanding of brain development constrain both modeling approaches and the validation of models against \textit{in vivo} brain development~\cite{khodosevich2023neurodevelopmental}.
Dendrite morphology, such as variations in arbor shape and size, is crucial in understanding the etiology of NDDs. 
Morphological irregularities in dendrites, essential for neural function and circuit connectivity, have been observed more frequently in children with certain NDDs~\cite{copf2016impairments, myers2019morphological}. 
These irregularities may disrupt neurite networks, highlighting the need for robust modeling and a deeper understanding of the developmental morphological transformations. 
Understanding the biophysical mechanisms governing these irregularities is essential for informing treatment development and guiding future research efforts.

Neuron morphological transformation, characterized by the branching and extension of dendrites and axons, is essential to neurite network functionality and neural circuit formation~\cite{liao2023semi}. 
Understanding these transformations is key to studying cell survival and validating computational models against experimental observations~\cite{liao2021quantification}. 
Often expensive and time-intensive, experimental studies highlight the need for efficient and fast computational approaches~\cite{isomura2023experimental, soriano2023neuronal}. 
Recent computational models have brought insights into neurodevelopmental processes, including neurite outgrowth~\cite{hentschel_instabilities_1994}, axon differentiation~\cite{krottje_mathematical_2007, pearson_mathematical_2011}, and guidance~\cite{aeschlimann_biophysical_2001}. 
While phenomenological models provide morphological insights~\cite{cuntz_one_2010, torben-nielsen_context-aware_2014}, they often neglect critical biophysical mechanisms~\cite{eberhard_neugen_2006, van_ooyen_independently_2014}. 
Incorporating these mechanisms introduces computational challenges, particularly for intracellular material transport~\cite{li_deep_2021, li2022modeling3D} and complex 3D neuronal structures~\cite{li2023isogeometric, li2022modeling}. 
Methods such as phase field approaches~\cite{takaki_phase_field_2015} and biophysically coupled models~\cite{qian_modeling_2022, qian2023biomimetic}, alongside isogeometric analysis (IGA) in biomedical fields~\cite{zhang_challenges_2013, zhang_geometric_2018, zhang2007patient}, enable accurate simulation of complex morphological transformations, potentially aligning experimental and computational neuroscience for target treatments.
For NDDs, computational models have shown promising results in simulating neurite deterioration~\cite{qian2025neurodevelopmental}, addressing the scarcity of experimental datasets and paving the way for a high-throughput data-driven approach.

Initially developed in the aerospace industry~\cite{singh2021digital}, digital twin refers to a virtual representation of a physical system connected through data flow, enabling advanced computational methods to improve physical system performance and decision-making process~\cite{jones2020characterising, tao2018digital}. 
While conventional simulations often focus on early-stage validation and optimization, they typically overlook runtime applications, leading to inefficiencies~\cite{liu2021review}. 
Integrating ML models into digital twin frameworks addresses these limitations, enabling dynamic analysis and real-time system optimization~\cite{rathore2021role}. 
Predicting neurite deterioration is a future frame prediction task for cell cultures, which is a challenging ML problem with broader applications in fields like autonomous driving and surveillance~\cite{oprea2020review}. 
Transformer-based architectures, such as VideoGPT and the Video Transformer Network, have shown success in capturing complex spatial and temporal dependencies through mechanisms like attention and vector quantization~\cite{yan2021videogpt, neimark2021video}. 
Efficient transformer models for video prediction, including PredFormer, further highlight the adaptability of transformers as robust spatiotemporal learners~\cite{ye2023video, tang2024predformer}. 
Building upon these advancements, MetaFormer has emerged as a strong candidate for spatiotemporal prediction tasks~\cite{yu2022metaformer, tan2023openstl}.
In this paper, we apply MetaFormer with tailored parameters to synthetic and experimental neurite deterioration datasets, effectively capturing the intricate spatiotemporal dynamics of neuronal development and degeneration.

This study tackles the task of neurite deterioration prediction by leveraging a MetaFormer architecture-based digital twin framework to achieve accurate and efficient long-term predictions of neurite dynamics. 
Facing difficulties associated with collecting experimental datasets, we incorporate an IGA-based phase field model as a synthetic data generator, enabling the generation of diverse biomimetic datasets. 
This work bridges the gaps between data availability, machine learning, and the complex nature of NDDs by combining the spatiotemporal machine learning model with synthetic data generation.
The main contributions of this paper are:
\begin{itemize}
    \item Developing a high-throughput end-to-end digital twin framework combining synthetic data generation, experimental images and morphology predictions, seamlessly integrating the simulation and analysis of neurite deterioration processes;
    \item Generating a novel synthetic dataset simulating diverse and realistic neurite deterioration patterns, overcoming the challenges posed by limited and expensive experimental datasets;
    \item Tailoring the MetaFormer-based gated spatiotemporal attention (gSTA) architecture to capture intricate temporal dependencies and morphological transformations, enabling accurate predictions of neurite dynamics;
    \item Applying a combined loss function with VGG16-based perceptual loss \cite{simonyan2014very} and mean square error (MSE), enhancing the model to capture subtle structural details in neurite images and improving prediction.
\end{itemize}

The rest of the paper is organized as follows: 
Section~\ref{sec: overview} provides an overview of the high-throughput digital twin framework for predicting neurite deterioration. 
Section~\ref{sec: NDDs review} reviews existing NDDs models and highlights their relevance to this work. 
Section~\ref{sec: ML model} illustrates the details of the neurite digital twin framework, including the data generation, methodologies, and ML models used. 
Section~\ref{sec: results} presents the results of the high-throughput digital twin framework, highlighting the performance of the ML model on both synthetic and experimental data.
Section~\ref{sec: conclusion} concludes the findings and discusses potential future research.

\section{Overview of high-throughput digital twin framework}
\label{sec: overview}

\begin{figure}[!htb]
\centering
\includegraphics[width=\textwidth]{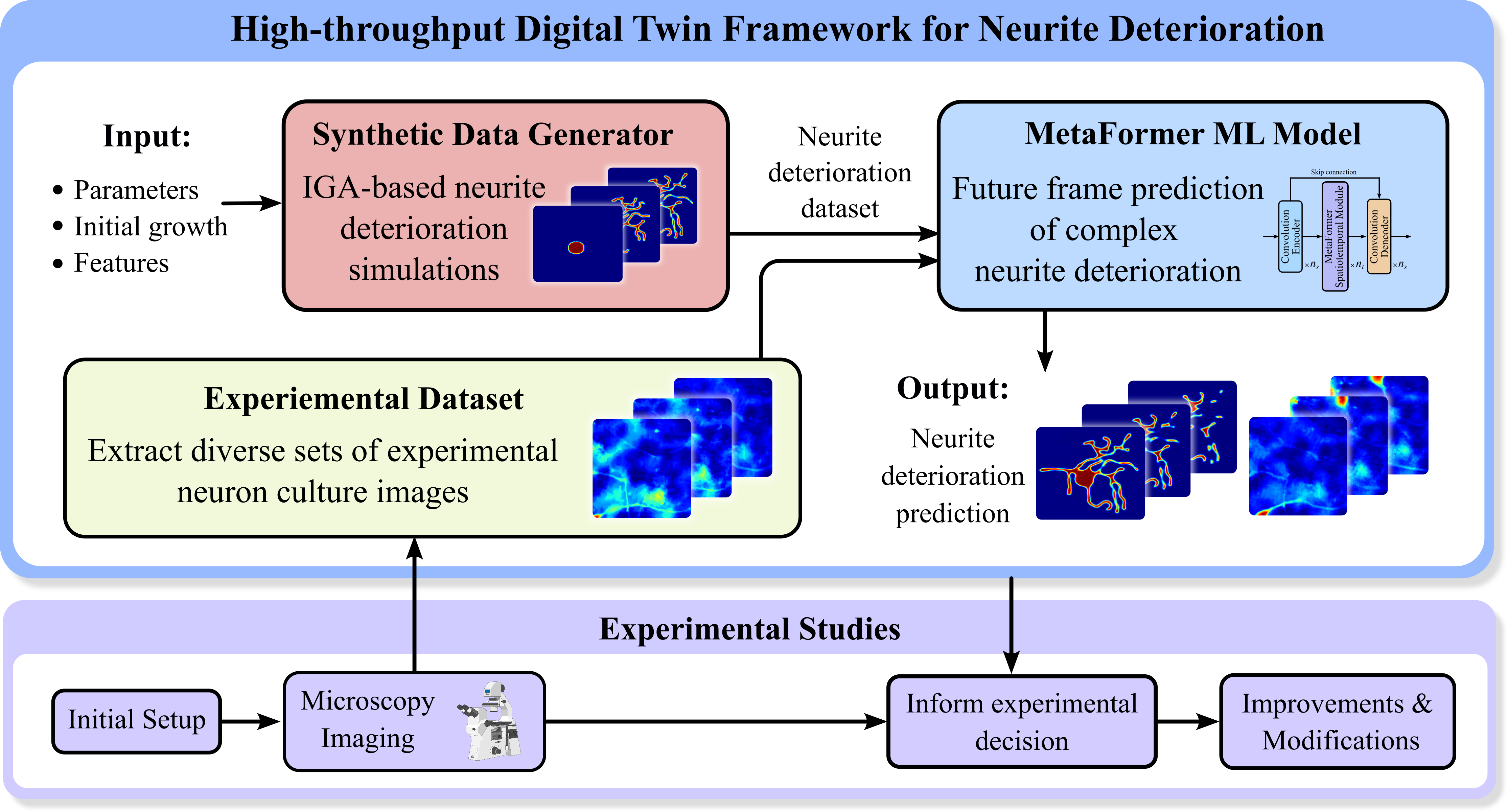}
    \caption{Overview of the proposed digital twin framework for predicting neurite deterioration. 
    The framework combines an IGA-based synthetic data generator (red module) and experimental neuron culture images (green module) to adapt the model to different scenarios. 
    The MetaFormer-based ML model (blue module) with gSTA provides accurate neurite deterioration predictions, and the experimental process (purple module) integrates real-world data to guide experimental design and decision-making.}
\label{fig:overview}
\end{figure}

The proposed digital twin framework (Figure~\ref{fig:overview}) comprises three modules that collectively address the complexities of neurite dynamics prediction: synthetic data generator (red module), experimental dataset (green module), and MetaFormer-based ML model (blue module). 

\begin{itemize}
    \item \textbf{The synthetic data generator} (red module) utilizes an IGA-based phase field model~\cite{qian2025neurodevelopmental} to simulate neurite deterioration, creating a comprehensive dataset that captures intricate deterioration morphological transformations with high precision. 
    The synthetic dataset provides a robust foundation for training ML models and offers a controlled setup for studying NDDs-induced deterioration.
    \item \textbf{The experimental dataset} (green module) consists of real-world neuron culture videos.
    This dataset ensures that the digital twin framework can generalize to diverse biological scenarios despite the limited availability of experimental data. 
    Combining synthetic and experimental datasets, the digital twin framework leverages the strengths of controlled simulations and real-world observations.
    \item \textbf{The MetaFormer-based ML model} (blue module) bridges the data-driven approach with advanced neural architectures. 
    The model accurately predicts future neurite growth and morphological transformation frames using the gSTA mechanism~\cite{gao2022simvp}. 
    This integration of synthetic data, experimental observations, and advanced ML techniques constructs a scalable and high-throughput pipeline capable of deepening our understanding of neurite deterioration and aiding the study of NDDs.
\end{itemize}

The experimental studies integrate microscopy images to guide experimental design and decision-making. They leverage the digital twin framework by capturing neurite morphological changes through microscopy imaging. The captured data is then fed into the framework, and the model generates predictions to aid experimental decisions. This feedback loop improves experiment setups, reduces costs, and accelerates research progress for neurite dynamics and NDDs-related behaviors.

\section{Review of IGA-based phase field NDDs model}
\label{sec: NDDs review}

NDDs severely impair motor and cognitive functions, with life-long lasting effects on patients. 
To understand the complex neurite deterioration processes associated with NDDs, characterized by retracting cellular boundaries and intricate morphological changes, IGA and phase field methods are combined to simulate the deterioration process in our earlier work~\cite{qian2025neurodevelopmental}.
These two advanced numerical techniques are well-suited for modeling complex boundary-changing phenomena in engineering and biological systems~\cite{gomez_accurate_2014, schillinger_isogeometric_2015}.
IGA is a high-order computational method that accurately represents smooth geometries without requiring the discretization typically associated with finite element methods~\cite{hughes_isogeometric_2005}. 
Meanwhile, the phase field method is particularly effective for capturing evolving boundaries, such as those observed in crack propagation and dendritic solidification, ideal for modeling dynamic neuron morphological transformations~\cite{takaki2014phase}.

The adopted IGA-based phase field model~\cite{qian_modeling_2022, qian2023biomimetic} incorporates five main equations. 
The phase field governing equation, which captures neuron morphological transformations, is defined as:
\begin{equation}
    \frac{\partial \phi}{\partial t} 
    = M_{\phi} \Bigg[ \nabla \cdot \left(a(\Psi)^2 \nabla \phi \right) 
    - \frac{\partial}{\partial x}\left(a(\Psi) \frac{\partial a(\Psi)}{\partial \Psi} \frac{\partial \phi}{\partial y}\right)
     + \frac{\partial}{\partial y}\left(a(\Psi) \frac{\partial a(\Psi)}{\partial \Psi} \frac{\partial \phi}{\partial x}\right)
     + \phi(1-\phi) \left(\phi - \frac{1}{2} + F_{driv} + 6H |\nabla \theta| \right) \Bigg], \label{eqn:phase_field_equation}
\end{equation}
where $\phi$ is the evolving phase field, $M_{\phi}$ is the mobility coefficient, $a(\Psi)$ is the anisotropy coefficient, $F_{driv}$ is the driving force for growth, $H$ is a constant value, and $\theta$ is the orientation term. 
The intracellular tubulin equation describing tubulin transport during neurite elongation is:
\begin{equation}
    \frac{\partial (\phi \,c_{tubu})}{\partial t} 
    = \delta_t \nabla\cdot (\phi \, \nabla c_{tubu}) 
    - \alpha_t \cdot \nabla (\phi \, c_{tubu})
    - \beta_{t} (\phi \, c_{tubu})
    + \epsilon_0 |\nabla(\phi_0)|^2 \int |\nabla(\phi_0)|^2 \, d \Omega, \label{eqn:tubulin_equation}
\end{equation}
where $c_{tubu}$ is the concentration of tubulin, $\delta_t$ is the diffusion rate, $\alpha_t$ is the active transport coefficient, $\beta_t$ is the decay coefficient, and $\epsilon_0 \frac{|\nabla(\phi_0)|^2}{\int{|\nabla(\phi_0)|^2} d\Omega}$ is the constant production. 
Here, $\phi_0$ is the initial phase field, and $\epsilon_0$ is the production coefficient. 
The competitive tubulin consumption equation, which models localized utilization of tubulin at neurite tips, is defined as:
\begin{equation}
    \frac{dL}{dt} = r_{g} \, c_{tubu} - s_{g}, 
    \label{eqn:dLdt_equation}
\end{equation}
where $\frac{dL}{dt}$ captures the dynamic tubulin consumption for neurite extension~\cite{mclean2004mathematical}, and $r_g$ and $s_g$ are the tubulin assembly and disassembly rates~\cite{mclean2004continuum, van_ooyen_competition_2001}. 
The synaptogenesis equation, which regulates neurotrophin particle dynamics and their degradation~\cite{qian2025neurodevelopmental}, is given as:
\begin{equation}
    \frac{\partial c_{neur}}{\partial t} 
     = D_c \nabla^2 c_{neur} 
    + (K - k_{p75}c_{neur})\frac{\partial \phi}{\partial t} \notag - k_2 c_{neur}, \label{eqn:synaptogenesis_equation}
\end{equation}
where $c_{neur}$ is the neurotrophin concentration, $D_c$ is the diffusion coefficient, and $K$ is the latent neurotrophin. 
$k_{p75}$ represents the degradation rate due to binding with $p75NTR$ receptors, while $k_2 c_{neur}$ acts as the sink term for neurotrophin degradation~\cite{nella2022bridging, krewson1996transport}. 
The driving force equation, integrating the effects of tubulin and neurotrophin, couples them back into the phase field~\cite{qian2025neurodevelopmental}:
\begin{equation}
    F_{driv} 
    = \frac{\alpha}{\pi} \tan^{-1}\left[H_\epsilon\left(\frac{dL}{dt}\right)\gamma (c_{opti}-c_{neur})\right], 
    \label{eqn:driving_force_equation}
\end{equation}
where $\frac{\alpha}{\pi}$ is a scaling coefficient, $H_\epsilon$ is the Heaviside step function, and $\gamma$ is the interfacial energy constant. 
The inverse relationship between neurotrophin and neuronal survival is incorporated by introducing the optimal neurotrophin concentration $c_{opti}$.
The phase field equation (Eqn.~\ref{eqn:phase_field_equation}) is driven by $F_{driv}$ (Eqn.~\ref{eqn:driving_force_equation}). 
We solve Eqns.~\ref{eqn:phase_field_equation}, \ref{eqn:tubulin_equation}, and \ref{eqn:synaptogenesis_equation} concurrently using IGA, coupled through Eqns.~\ref{eqn:driving_force_equation} and \ref{eqn:dLdt_equation}.
For a detailed explanation of the IGA-based phase field model, we refer readers to our earlier work~\cite{qian2025neurodevelopmental}.

We implement the model on locally refined truncated T-splines using IGA to effectively model intricate morphological characteristics of neurite structures efficiently~\cite{hughes_isogeometric_2005,wei2017truncated,liunurbs}.Unlike Non-Uniform Rational B-Splines and B-splines~\cite{piegl2012nurbs}, which lack local refinement capabilities, T-splines incorporate T-junctions, allowing for localized refinements that reduce computational degrees of freedom while maintaining exact geometric representations~\cite{wei2015truncated,wei2016extended,wei2022analysis}. 
This is particularly desirable for capturing the neurite morphological complexities of NDDs.
The model was written in C++ using the PETSc library~\cite{petsc-user-ref, petscsf2022} with mesh paritioning~\cite{karypis1998fast} for enhanced computational efficiency and scalability~\cite{qian_modeling_2022}.
In this paper, we aim to leverage the success of combining data-driven techniques and IGA numerical methods in the computational neuron modeling field~\cite{li2023isogeometric,li_reaction_2020,li_deep_2021}.

\section{Neurite deterioration digital twin using MetaFormer-based gSTA}
\label{sec: ML model}

\begin{figure}[ht]
\centering
\includegraphics[width=\textwidth]{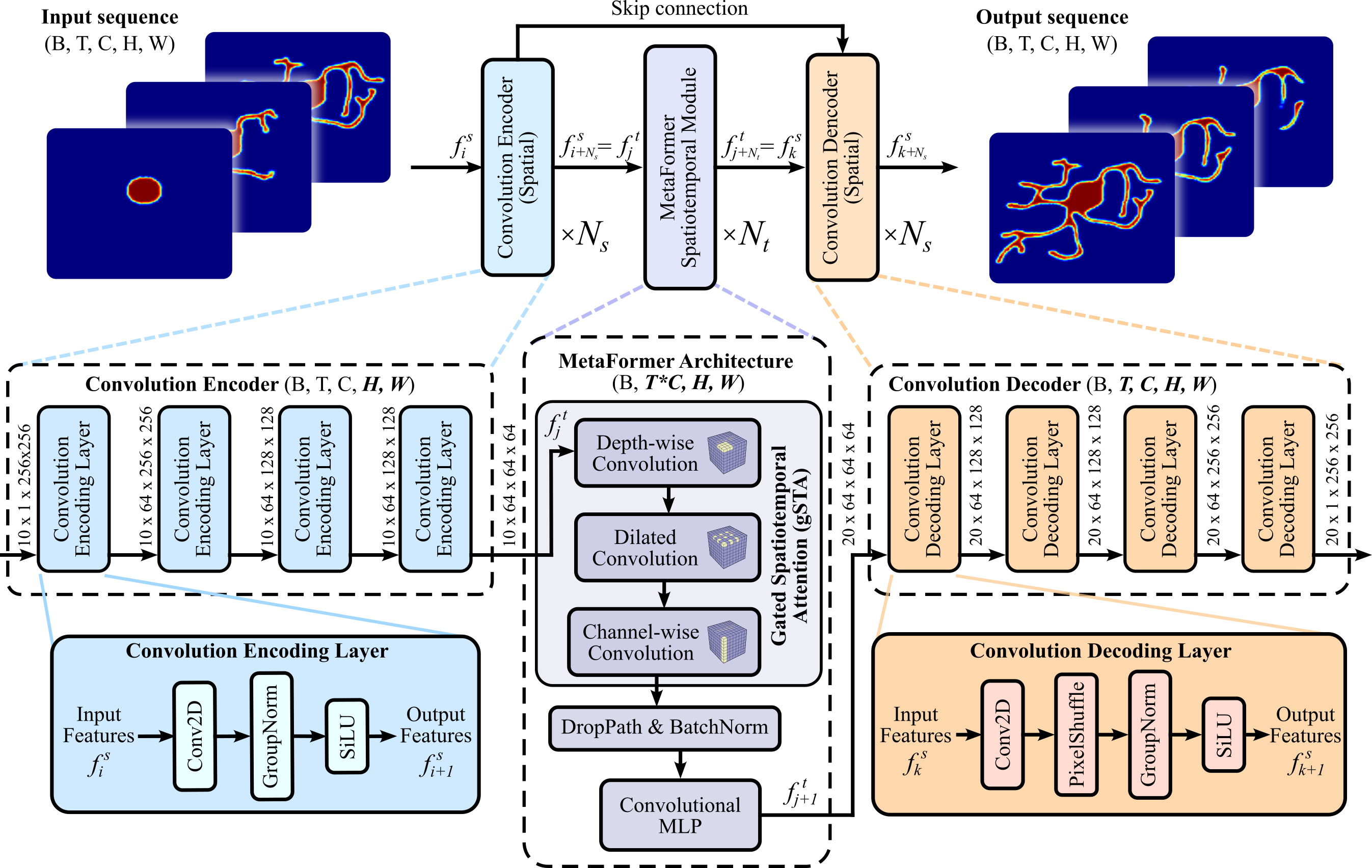}
    \caption{MetaFormer-based architecture overview. 
    The proposed model predicts future neurite growth frames from input sequences via a spatial encoder, spatiotemporal extractor, and spatial decoder. 
    The encoder compresses spatial dimensions to a latent representation. 
    Repeated multiple times, the spatiotemporal extractor combines depth-wise and dilated convolutions with MLPs. 
    The decoder upsamples features to reconstruct output frames, effectively modeling spatial-temporal dynamics of neurite growth.}
\label{fig:model_structure}
\end{figure}

Predicting neurite deterioration is complex due to its multi-stage, diverse morphological transformations and intricate thin neurite structures. 
Computational biophysics-based solvers, while capturing fine-grained morphological details and exhibiting good computational efficiency uplift over past methods~\cite{takaki2014phase,qian2025neurodevelopmental}, are still computationally expensive. 
High-fidelity simulation often involves solving a system of coupled non-linear equations over complex geometries. 
This leads to significant memory and computational costs, especially for large-scale neurite networks, calling for ML-based approaches to leverage the immense parallelization power of GPUs.
Previous convolutional-based ML surrogate models have demonstrated good neurite behavior prediction~\cite{qian2023biomimetic}, but they are inherently limited to single-step predictions, falling short of capturing sequential relationships and leveraging temporal dependencies inherent in neurite dynamics. 
As a result, these approaches often produce independent predictions, lacking coherence across successive neurite behavior changes. 
Additionally, convolutional recurrent methods, which rely on step-by-step predictions, suffer from cumulative error propagation~\cite{shi2015convolutional}, leading to significant performance degradation over extended time frames.

To address these challenges, we adopt the SimVP model equipped with the gSTA module~\cite{gao2022simvp, tan2022simvp}, a MetaFormer-based architecture designed for spatiotemporal predictions (Figure \ref{fig:model_structure}). 
Using the existing library from OpenSTL\cite{tan2023openstl}, we configure the model to suit the specifics of neurite morphological transformation prediction.
This model processes input sequences with dimensions (batch, time, channels, height, width) (B, T, C, H, W), as detailed in our data generation section (Section \ref{sec:datagen}). 
By integrating spatial and temporal features in a unified end-to-end pipeline, our approach overcomes the limitations of previous CNNs-based approaches~\cite{qian2023biomimetic}, enabling more accurate predictions of complex neurite deterioration processes.

\subsection{Novel synthetic deterioration and experimental dataset generation}
\label{sec:datagen}

Experimental neuron culturing is expensive and time-intensive. 
It requires careful planning, specialized equipment, and precise environmental controls, such as sterility and nutrition for cell survival~\cite{liao2023semi}. 
These aspects of the experiment make data collection very challenging and impractical for fast and short-term ML model development. 
To address this issue, we employ an IGA-based phase field model as a synthetic data generator, enabling the generation of a diverse neurite morphology dataset that covers various growth patterns and deterioration behaviors.

\textbf{Synthetic dataset.} 
Our data processing workflow begins by extracting VTK files from each simulation time series. 
We run the IGA-based phase field model to generate 134 simulation cases.
Each simulation is initialized with a random orientation variable to introduce different neurite behaviors.
Neurite retraction behaviors are simulated by setting an initial $c_{opti}$ value of 1, progressively reduced to 0 at 150,000 iterations. 
This reflects the increasing $c_{neur}$ magnitude and its inverse relationship with neuron survival. 
The simulations are conducted on an unstructured truncated T-spline mesh, which requires specialized preprocessing for accurate data handling~\cite{wei2017truncated}.
We utilize a cKD tree algorithm to optimize this process for efficient nearest-neighbor searches and spatial interpolation, significantly reducing computational overhead and enhancing preprocessing speed~\cite{bentley1975multidimensional}.
We evenly sample the first 10 input frames from the initial 150,000 iterations and the next 20 frames for prediction from the subsequent simulation steps, carefully excluding any cases with insufficient or corrupted data. 
To ensure consistency across all cases, we standardized the domain size to $256 \times 256$ pixels, determined by the global minimum and maximum coordinates across all simulations. 
Each frame is interpolated onto a structured grid within these bounds using the cKD tree for optimized spatial mapping. 
Fallbacks are implemented to handle missing or corrupted files, such as substituting with recent valid data when necessary. 
This streamlined workflow ensures efficient edge-case handling for a robust analysis-ready dataset.

\textbf{Experimental dataset.}
Human iPSC-derived neurons are cultured over several days, with images captured at 10-minute intervals for 96 hours.
These time-lapse images are compiled into videos to capture the morphological transformations of the neuron cultures over time.
For detailed documentation of the experimental protocols, we refer readers to~\cite{zhao2017apoe,kawatani2023abca7}.
Since collecting an experimental dataset is time-intensive and costly, we preprocess 10 neuron culture videos into the experimental dataset.
Each video, approximately 11 seconds long at 25 frames per second, is segmented into nine sequential segments to ensure a reasonable movement within each time step, resulting in nine samples per video. 
Each generated sample consists of 30 frames, with the first 10 frames serving as input and the subsequent 20 frames designated as output ground truth for comparison with predictions. 
To align with the input requirements of our ML model, each frame in a sample is also divided into 16 smaller sections (a $4 \times 4$ split), where each section measures $256 \times 256$ pixels. 
This strategy preserves spatial granularity, increasing the dataset size while saving critical features for accurate model training without downsampling. 
By applying consistent preprocessing standards across all samples, we can ensure the robustness and suitability of the dataset for prediction tasks.

\subsection{Convolution-based encoding and decoding}
\label{subsec:ML architecture}

The model (Figure~\ref{fig:model_structure}) employs a CNN autoencoder structure to process complex spatiotemporal data effectively.
The encoder extracts spatial features from input image sequences through convolutional layers.
The convolutional encoding layers employ 2D convolutions, GroupNorm, and SiLU activation to extract spatial features 
$f^s_{i}$ efficiently, defined as:
\begin{equation}
\begin{aligned}
f^s_{i} = \sigma(GN(\text{Conv2D}(f^s_{i-1}))), \hspace{5mm}
1 \leq i \leq N_{s},
\end{aligned}
\end{equation}
where $N_s$ is the number of encoding layers ($N_s = 4$ in our study), and $\sigma$ is the Sigmoid Linear Units (SiLU) activation function~\cite{elfwing2018sigmoid, nwankpa2018activation}, which has shown better performance in modern deep ML models.
$GN$ is the group normalization.
The spatial features are progressively extracted through multiple encoding layers to capture the intricate morphological patterns in the input data.
The decoder reconstructs the output predictions from the encoded feature representations.
It uses convolutions and upsamples with a PixelShuffle~\cite{shi2016real} to enhance computational efficiency and spatial resolution.
The spatial decoding feature, $f^s_{k}$ is defined as:
\begin{equation}
\begin{aligned}
f^s_{k} = \text{PixelShuffle}(\sigma(GN(\text{Conv2D}(f^s_{k-1})))), \hspace{5mm}
N_s + N_t < k \leq 2N_s + N_t,
\end{aligned}
\end{equation}
where $N_t$ is the number of temporal layers ($N_t = 16$ in our study), which will be discussed in Section~\ref{subsec:metaFormer}.
The range of $k$ ensures the decoder matches the encoder in the number of layers, enabling correct reconstruction of spatial features after processing the temporal layers.
This design enables the model to learn spatial and temporal dependencies effectively, ensuring accurate reconstruction of the neurite deterioration across time steps.

\subsection{MetaFormer architecture with gated spatiotemporal attention mechanism}
\label{subsec:metaFormer}

Figure~\ref{fig:model_structure} showcases the detailed machine learning model layers and structures.
We utilize MetaFormer architecture (purple module) as the backbone for our temporal translator~\cite{yu2022metaformer}.
The MetaFormer framework generalizes the Transformer design~\cite{vaswani2017attention} by changing the token-mixing module, which can be replaced with gSTA mechanism~\cite{gao2022simvp}. 
Unlike conventional self-attention, gSTA focuses on efficient spatial and temporal learning through localized and dilated convolutions, lowering computational costs while maintaining effective attention performance.
This modification allows for the efficient processing of spatiotemporal data, which is essential for predicting neurite morphological transformations. 
Drop path regularization and batch normalization enhance model robustness, helping stabilize training and improve generalization across diverse datasets. 
The convolutional multilayer perceptron (MLP) with activation function is placed after the token-mixing module to introduce non-linearity into the model (Figure~\ref{fig:model_structure}).
The architecture is designed to effectively capture intricate temporal patterns and evolving morphological structures inherent in neurite deterioration, providing a scalable and efficient architecture for handling complex spatiotemporal dynamics.

\begin{figure}[ht]
\centering
\includegraphics[width=0.65\columnwidth]{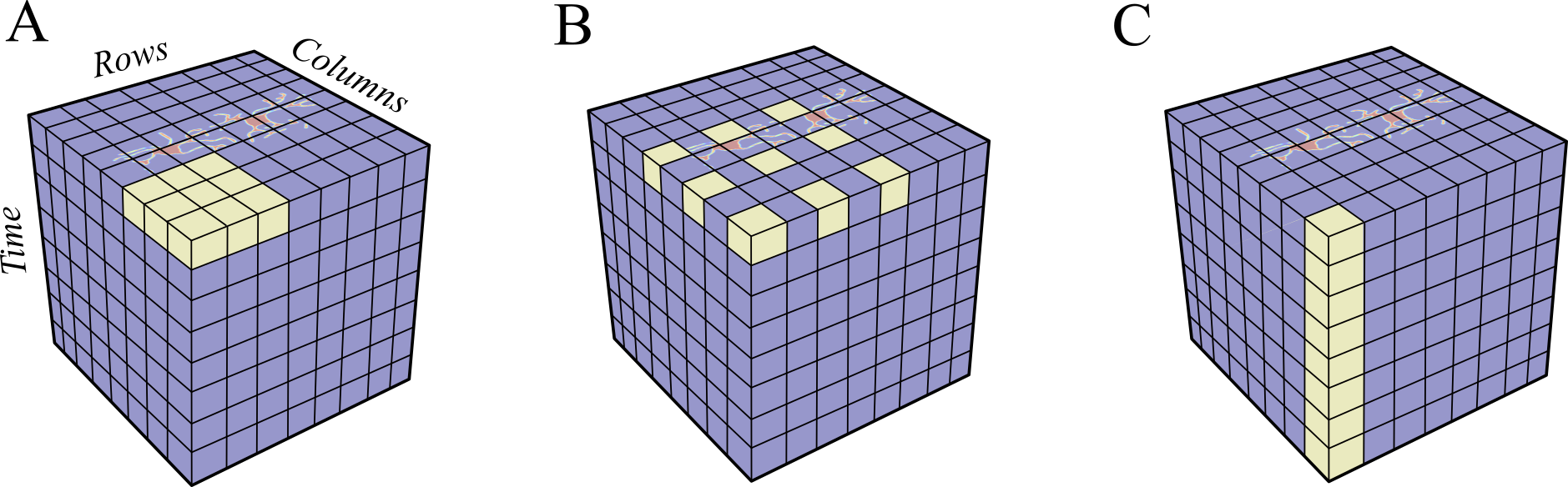}
\caption{gSTA for learning temporal characteristics \cite{tan2022simvp}. 
(A) Depth-wise convolution convolution to collect information from the local reception field.
(B) Dilated convolution to obtain relatively distant pixel information.
(C) Channel-wise convolution is applied across time dimensions to capture temporal features.
}\label{fig:gSTA}
\end{figure}

We incorporate the gSTA module into the MetaFormer architecture~\cite{tan2022simvp} to better model the evolving morphological patterns within our neurite growth dataset. 
By explicitly focusing on regions of interest across both spatial and temporal dimensions, the gSTA module prioritizes critical morphological transformation, such as changes in neurite thickness, branching, and overall structure, while suppressing less informative background noises. 
As it transforms the initial 10 input frames into 20 predicted frames, the attention mechanism in the module emphasizes predominant spatiotemporal patterns, effectively capturing subtle neurite deterioration trends that emerge over time. 
Combining a series of spatiotemporal translators (Figure~\ref{fig:gSTA}), the gSTA module ensures that the resulting representation is attuned to the complex temporal evolution characteristic of neuronal processes.
The spatiotemporal translator units include:
\begin{itemize}
    \item \textbf{Depth-wise convolution (ConvDW)}: Focusing on spatial features by capturing local receptive fields within individual channels.
    \item \textbf{Dilated convolution (ConvDW-d)}: Expanding the receptive field across the spatial dimension, effectively connecting distant regions to model spatial features.
    \item \textbf{Channel-wise convolution ($\text{Conv2D}_{1 \times 1}$)}: Functioning similar to a gating mechanism in LSTM, capturing temporal dependence while dynamically modulating the flow of information by evaluating feature relevance.
\end{itemize}

The spatiotemporal feature, $f^t_j$, encodes a high-level representation of the neurites spatiotemporal information in the input images, combining both spatial morphological details and their temporal evolutions. 
Specifically, $f^t_j$ captures the shape, thickness, and branching patterns of neurites and how these characteristics change across consecutive frames. 
By focusing on the most relevant transformations and suppressing minor, less informative variations, $f^t_j$ provides a context-rich embedding of neurite morphology. 
This enables the model to accurately anticipate future growth patterns and identify subtle temporal trends within the neuronal dataset.
$f^t_j$ is defined as~\cite{gao2022simvp}:
\begin{gather}
    \hat f^t_j = \text{Conv2D}_{1 \times 1}(\text{ConvDW-d}(\text{ConvDW}(f^t_j))), \label{eq:conv_eq} \\
    \{g, \bar f^t_j\} = \text{split}(\hat f^t_j), \label{eq:split_eq} \\
    f^t_{j+1} = \sigma(g) \circ \bar f^t_j. \label{eq:combine_eq}
\end{gather}
where  $f^t_j$ represents the input feature map at layer $j$.
The $split$ operation splits the tensor across channel dimension into $g$ and $\bar f^t_j$, in which $g$ is then used with $\sigma$ to act as a gating mechanism.
$\sigma$  denotes the activation function, specifically the Gaussian Error Linear Unit (GELU)~\cite{hendrycks2016gaussian}, defined as $x\circ \Phi(x)$  where  $\Phi(x)$ is the standard Gaussian cumulative distribution function.
``$\circ$'' is element-wise multiplication.

This approach enables the model to extract and integrate spatial and temporal information efficiently, improving its ability to predict complex spatiotemporal patterns.
By extracting spatiotemporal attention from the latent space between the encoder and decoder, the gSTA module effectively captures temporal characteristics crucial for accurate video prediction. 
Using large kernel convolutions imitates attention mechanisms, enabling the network to build global context by sampling pixels beyond local regions. 
The gating mechanism dynamically adjusts weights based on the significance of spatial and temporal features, enabling the model to capture intricate dependencies across both dimensions, which aligns with the complex spatiotemporal relationships inherent in neurite deterioration.
Comparing the baseline CNN autoencoder for neurite behaviors~\cite{qian2023biomimetic}, while the baseline performs adequately for spatial feature extraction and predictions, the improved model achieves superior temporal learning and representation, enabling the prediction of complex neurite morphology evolutions over time and leverage morphological transformation information across time steps.

\subsection{Enhanced layers tailored for neurite deterioration predictions}
\label{subsec:layers_structures}

We configure the convolution encoding-decoding structure and the MetaFormer architecture with $N_S = 4$ spatial layers and $N_T = 16$ temporal layers (Figure~\ref{fig:model_structure}). 
4 layers of encoding and decoding provide sufficient channel depth to extract spatial features proven by the existing model~\cite{qian2023biomimetic}.
To balance computational cost with model complexity, we set the hidden dimensions to $hid_S=64$ for spatial features and $hid_T=256$ for temporal modeling, where $hid_S$ defines the spatial embedding dimensionality within each frame, and $hid_T$ defines the temporal embedding dimensionality across consecutive frames.
\begin{itemize}
    \item \textbf{Encoding layers}. The encoder processes the input tensor of size $(10 \times 1 \times 256 \times 256)$, corresponding to (time, channels, height, width) in a sample, using a series of convolutional layers (Figure~\ref{fig:model_structure} blue module).
    Starting with the input, the encoding layer transforms it to $(10 \times 64 \times 256 \times 256)$ through convolution while maintaining the spatial resolution. 
    Then, a stride of 2 is selected to reduce the spatial dimensions to $(10 \times 64 \times 128 \times 128)$. 
    Another convolutional layer keeps the size at $(10 \times 64 \times 128 \times 128)$, followed with a final downsampling step (stride = 2), outputing the tensor of size $(10 \times 64 \times 64 \times 64)$.
    \item \textbf{MetaFormer layers}. The tensor is then passed to the MetaFormer gSTA module, which increases the temporal dimension from 10 to 20 frames without affecting spatial size over 16 layers, outputting $(20 \times 64 \times 64 \times 64)$ (Figure~\ref{fig:model_structure} purple module). 
    Setting to 16 temporal layers enables the model to capture long-range dependencies and subtle temporal variations essential for evolving neurite dynamics.
    \item \textbf{Decoding layers}. The decoder then reconstructs the features by progressively upsampling the tensor using convolution and pixel shuffle operations (Figure~\ref{fig:model_structure} orange module). 
    Starting from $(20 \times 64 \times 64 \times 64)$, the decoding layer upsamples with pixel shuffle to a size of $(20 \times 64 \times 128 \times 128)$, followed by another convolutional layer maintaining the size at $(20 \times 64 \times 128 \times 128)$. 
    A subsequent pixel shuffle operation further upsamples the tensor to $(20 \times 64 \times 256 \times 256)$, and a final convolutional layer outputs the reconstructed features of size $(20 \times 1 \times 256 \times 256)$, matching the expected 20 frame prediction of neurite patterns.
\end{itemize}

This design enables the model to process high-resolution spatiotemporal data efficiently while maintaining adaptability to diverse datasets.
To further improve its applicability, we implement a customized input pipeline that processes sequences of 10 healthy growth frames as input and predicts the subsequent 20 frames of neurite deterioration. 
This setup tailors the SimVP architecture to meet the unique challenges of neurite morphology modeling. 
By introducing domain-specific architectural refinements, we extend the temporal prediction capabilities of SimVP and enhance its robustness. 
These advancements enable the model to capture complex, multi-stage neurite deterioration effectively, substantially improving spatiotemporal modeling for neuroscience applications.

\subsection{Combined MSE and perceptual loss function}
\label{subsec:loss_function}

Neurites are characterized by thin and intricate patterns, and pixel-wise loss functions like MSE often fail to capture visual similarity effectively.
This is because MSE is sensitive to small spatial shifts, resulting in high loss values despite similar neurite patterns.
We employ a combined loss function that integrates MSE with perceptual loss to overcome this limitation. 
Perceptual loss evaluates high-level features, such as edges, textures, and patterns, from the predicted and ground-truth images, making it invariant to minor spatial misalignments. 
Perceptual loss is derived by passing the prediction and ground-truth through a pre-trained convolutional neural network, such as VGG16~\cite{simonyan2014very}, which extracts meaningful high-level features from its intermediate layers. 
The combined loss function is:
\begin{align}
\mathcal{L}_{\text{total}}(\phi_{\text{pred}}, \phi_{\text{true}}) = \mathcal{L}_{\text{MSE}}(\phi_{\text{pred}}, \phi_{\text{true}}) 
+ \lambda \sum_{l} \left\| V_l(\phi_{\text{pred}}) - V_l(\phi_{\text{true}}) \right\|_2^2, \label{eqn:combined_loss}
\end{align}
where $\mathcal{L}_{\text{MSE}}$ is the \textit{MSE} between the predicted image $\phi_{\text{pred}}$ and the ground-truth image $\phi_{\text{true}}$.
$\lambda$ is the weighted coefficient of perceptual loss.
$V_l$ represents the $l$-th feature map extracted from the pre-trained VGG16 model.
$\left\| \cdot \right\|_2^2$ denotes the squared Euclidean distance between feature maps.
By combining perceptual loss with MSE, the model reduces pixel-wise discrepancies while capturing essential structural and textural features. 
This enables accurate and visually coherent predictions of neurite dynamics.

\section{Results}
\label{sec: results}

In this section, we showcase how our digital twin framework effectively predicts neurite deterioration across synthetic and experimental datasets. 
The synthetic dataset provides controlled settings to evaluate the capability of our model to capture complex temporal and morphological transformations in single and multiple neurite deteriorations. 
It establishes a robust baseline for testing predictive accuracy and temporal coherence.
Moving beyond synthetic settings, we then evaluate experimental neuron culture images. 
We showcase the capability of our digital twin framework to generalize to real-world biological data, highlighting its adaptability to diverse and complex neurite transformations. 
The results from synthetic and experimental datasets show the great potential for connecting simulation-based predictions with practical biological applications.

The dataset is divided into three subsets for ML training: 70\% is allocated to the training set, 15\% to the validation set, and the remaining 15\% to the test set.
The model takes the first ten healthy growth images from a simulation case as inputs and predicts the next 20 frames of neurite deterioration.
The input data dimensions \([10, 1, 256, 256]\) represent 10 temporal frames of high-resolution spatial features, aligning with the demands of accurate spatiotemporal prediction.
Training settings include a warmup learning rate of \(1 \times 10^{-6}\) for the initial five epochs, a one-cycle learning rate scheduler with a decay rate of 0.1 after 100 epochs, and a final divisor of \(1 \times 10^{4}\). 
The model is trained for 250 epochs with a batch size of 16, which varies depending on the specific GPU in use, and an initial learning rate \(lr = 0.001\).
The training in this paper is done using Nvidia Tesla V100:32 GPUs at Pittsburgh Supercomputer Center~\cite{ecss,xsede}.
Training requires approximately 8 hours, while inference for predicting each sequence is completed in a fraction of a second.

\subsection{ML prediction based on synthetic dataset}
\label{subsec:ML_prediction}

\begin{figure}[ht]
    \centering
\begin{subfigure}{0.49\textwidth}
    \centering
    \vtop{\null \hbox{\includegraphics[width=\textwidth]{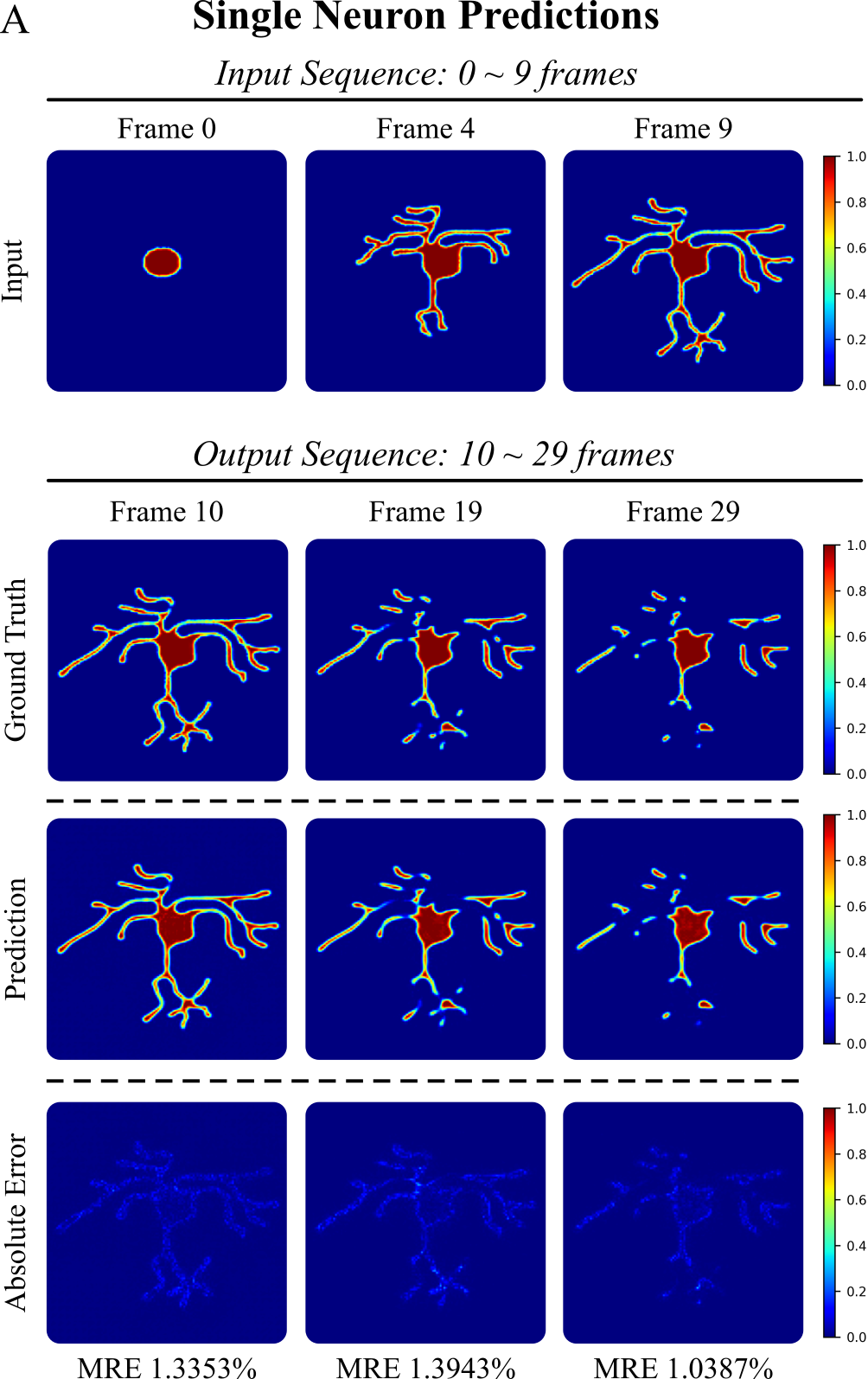}}}
\end{subfigure}
\hfill
\begin{subfigure}{0.49\textwidth}
    \centering
    \vtop{\null \hbox{\includegraphics[width=\textwidth]{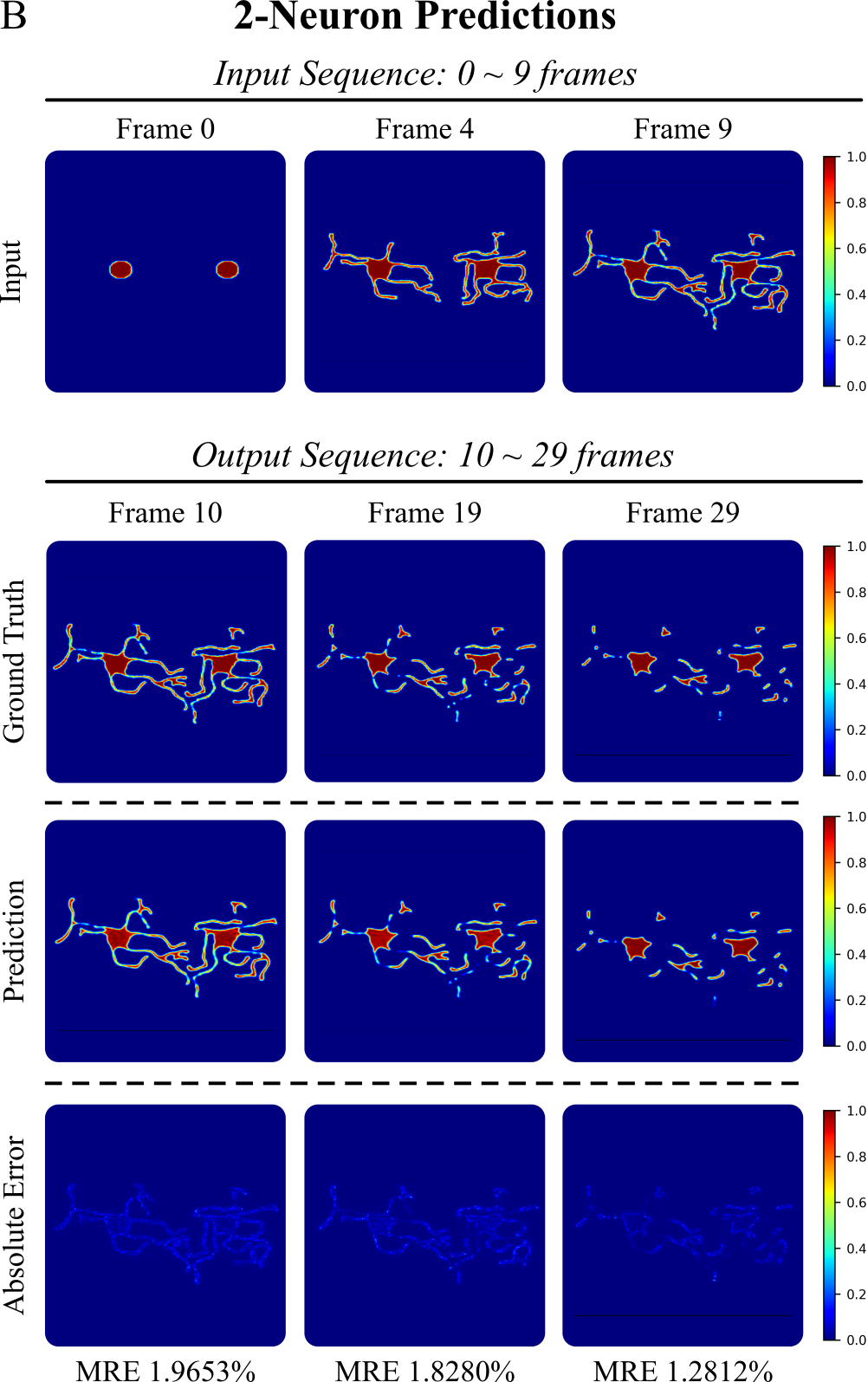}}}
\end{subfigure}
    \caption{(A\&B) Prediction for single and 2-neuron synthetic neurite deterioration. 
    The input sequence (top row) consists of frames from time steps 0–9, and the ground truth (second row) covers time steps 10–29. 
    The predicted sequence (third row) demonstrates that the model can predict future neurite deterioration. 
    Absolute error maps (fourth row) highlight pixel-wise discrepancies, with MRE percentages shown for sampled frames.}
    \label{fig:synthetic_predictions_1}
\end{figure}

\begin{figure}[ht]
    \centering
\begin{subfigure}{0.49\textwidth}
    \centering
    \vtop{\null \hbox{\includegraphics[width=\textwidth]{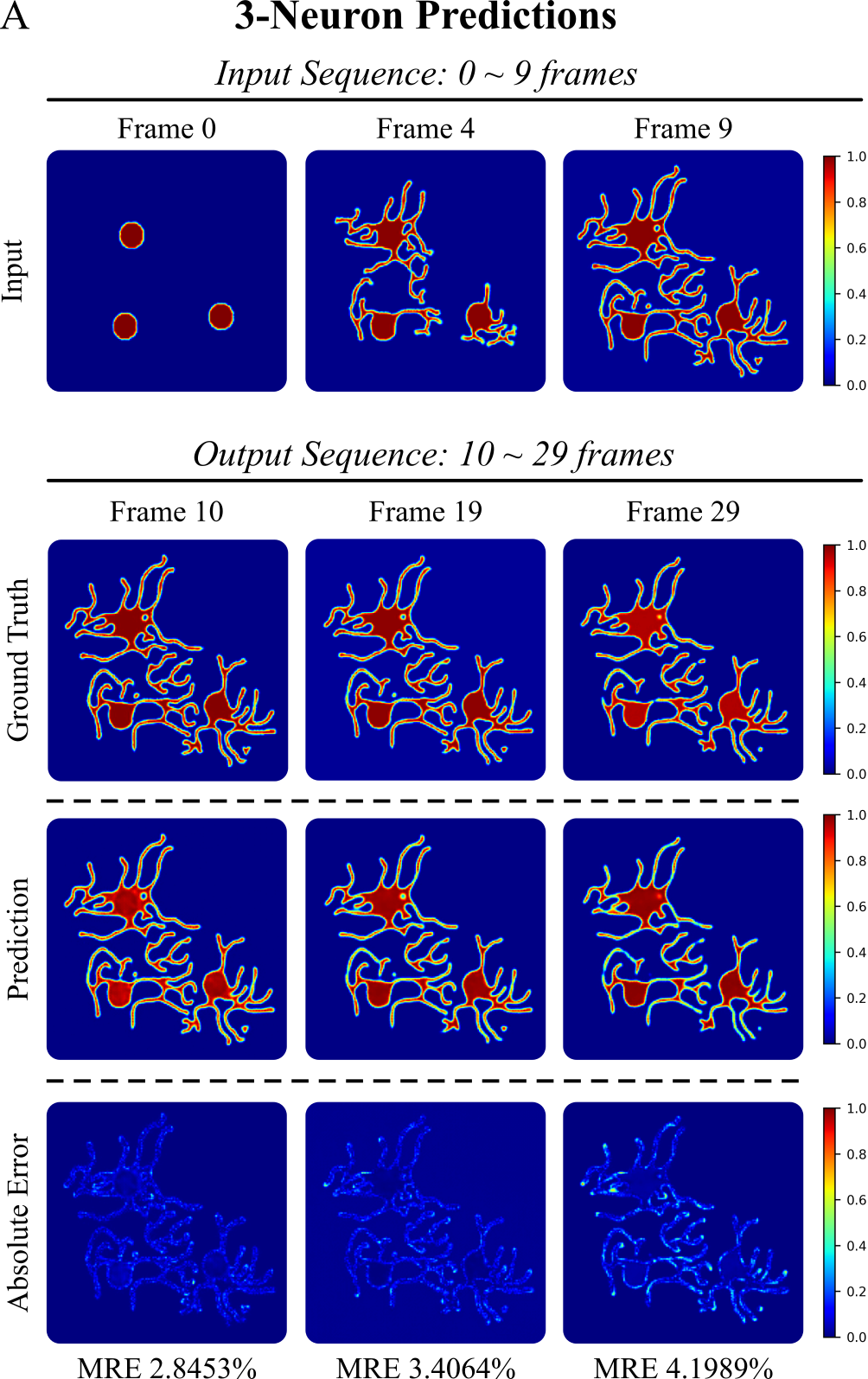}}}
\end{subfigure}
\hfill
\begin{subfigure}{0.49\textwidth}
    \centering
    \vtop{\null \hbox{\includegraphics[width=\textwidth]{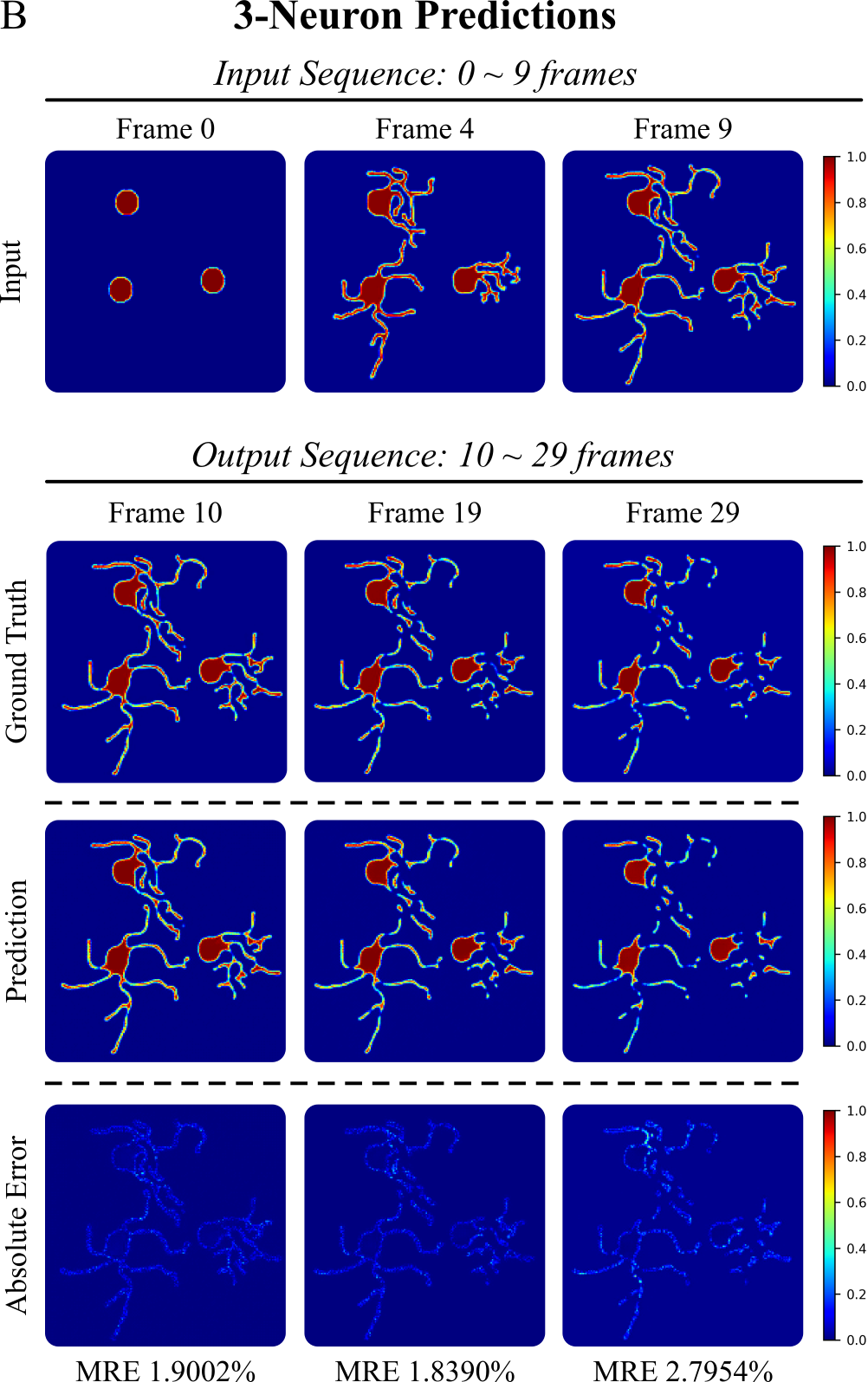}}}
\end{subfigure}
    \caption{(A\&B) Prediction for 3-neuron synthetic neurite deterioration. 
    The layout follows the same structure as Figure~\ref{fig:synthetic_predictions_1}.}
    \label{fig:synthetic_predictions_2}
\end{figure}

Figure~\ref{fig:synthetic_predictions_1} shows the prediction results of single and 2-neuron synthetic neurite deterioration, and Figure~\ref{fig:synthetic_predictions_2} shows the prediction results of 3-neuron synthetic neurite deterioration.
Each figure has the input sequence (top row) that spans frames from time steps 0–9, while the ground truth sequence (second row) represents the target output for time steps 10–29. 
The predicted sequence (third row) demonstrates the capability of the model to accurately predict the future spatiotemporal transformation of neurite deterioration, with errors fluctuating around 1 to 2\%.
Absolute error maps (fourth row) provide pixel-wise error visualizations, with Mean Relative Error (MRE) values quantified for keyframes~\cite{qian2023biomimetic, li2023isogeometric} because it provides a dimensionless measure of error, expressed as a percentage, making it easier to interpret and compare. 
The results showcase the effectiveness of our digital twin framework in capturing dynamic behaviors, as indicated by the low MRE values across predictions.

\subsection{ML prediction based on experimental dataset}

\begin{figure}[ht]
    \centering
\begin{subfigure}{0.49\textwidth}
    \centering
    \vtop{\null \hbox{\includegraphics[width=\textwidth]{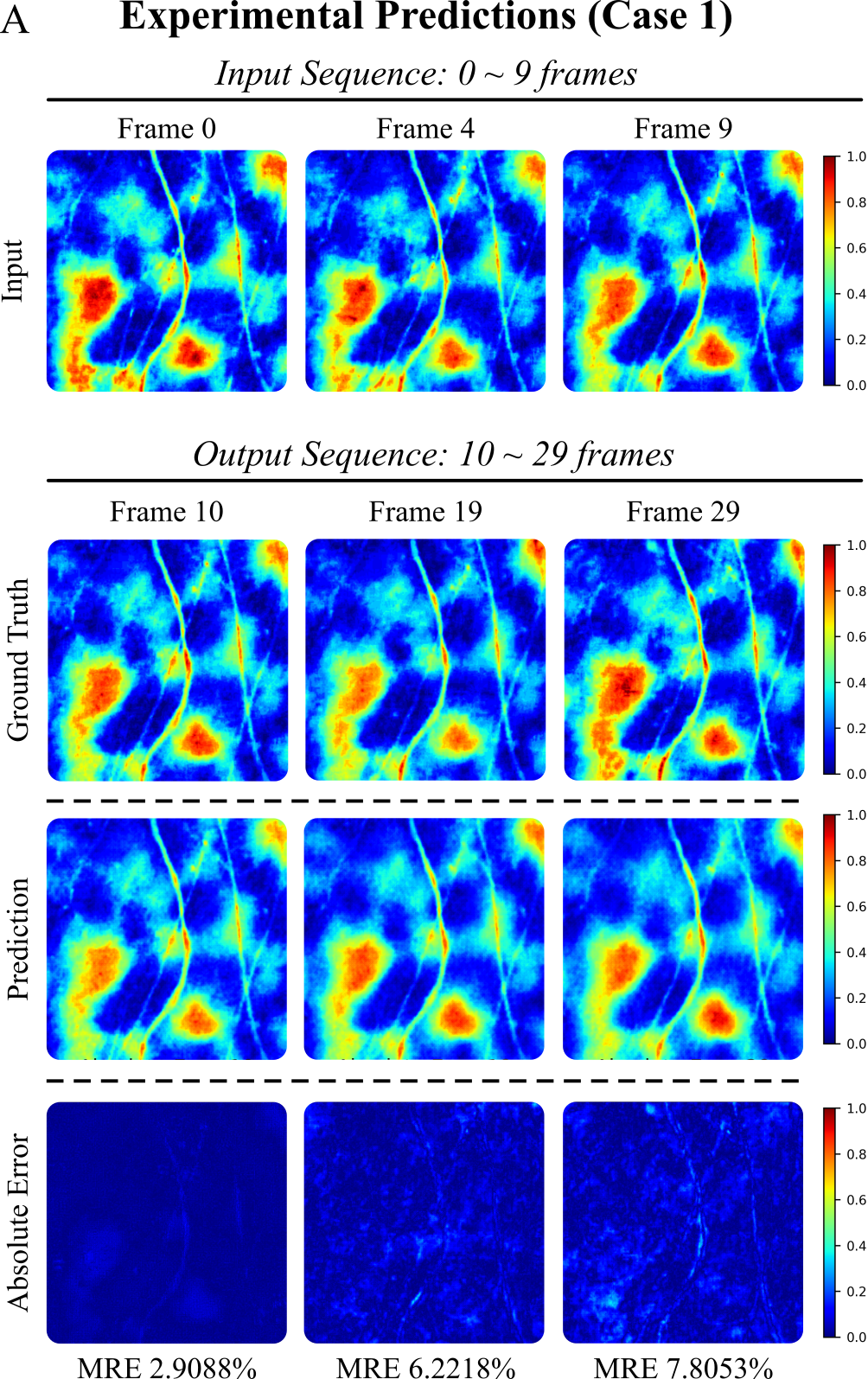}}}
\end{subfigure}
\hfill
\begin{subfigure}{0.49\textwidth}
    \centering
    \vtop{\null \hbox{\includegraphics[width=\textwidth]{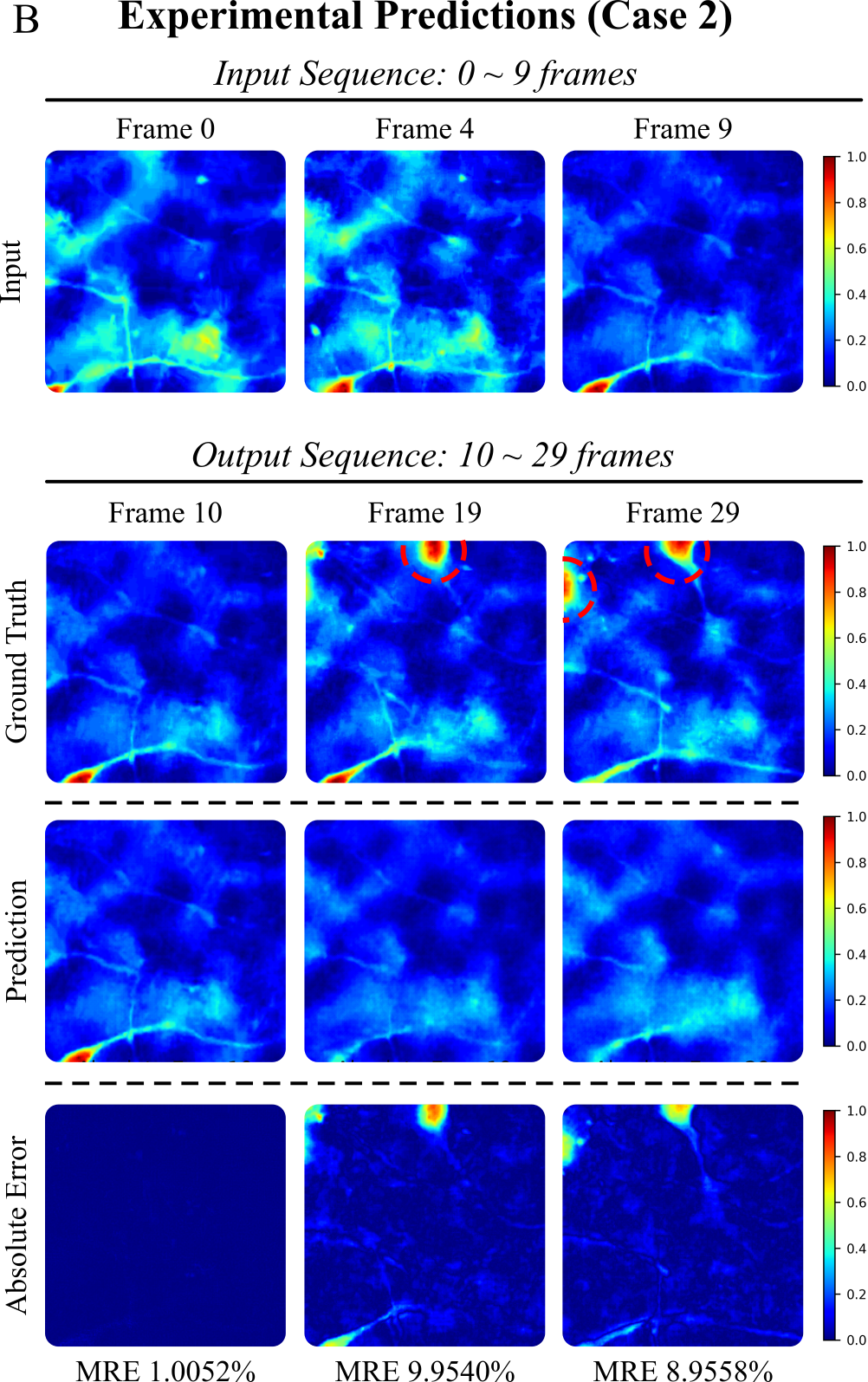}}}
\end{subfigure}
\caption{
    (A\&B): Predictions of experimental neurite transformations using a model trained on the experimental dataset.
    The layout follows the same structure as Figure~\ref{fig:synthetic_predictions_1}.
    Red dashed circle: unexpected neurons enter the frame, which the ML model fails to predict.}
    \label{fig:experimental_neuron_predictions}
\end{figure}

Figure~\ref{fig:experimental_neuron_predictions} illustrates the model predictions for experimental neurite deterioration cases. 
Similarly, the input sequence (top row) consists of time steps 0–9, and the ground truth sequence (second row) represents the target output for time steps 10–29. 
The predicted sequence (third row) demonstrates that the model can be generalized to real experimental datasets. 
Absolute error maps (fourth row) highlight pixel-wise discrepancies. MRE values are also quantified for keyframes, and errors fluctuate between 2\% and 4\%.
Compared to synthetic predictions, experimental predictions show relatively lower accuracy and performance due to added noise and gradient variations introduced by differing background concentrations and inconsistencies in microscope lenses.
The experimental dataset introduces significant challenges due to concurrent translation, scaling, and deformation variations, while convolution operations are primarily suited for handling small translational variance. 
Notably, red dashed circles in the error maps highlight unexpected neurons entering the frame, which the model fails to predict accurately. 
Despite these challenges, the results exhibit low MRE, showcasing the robustness of our model in predicting neurite dynamics.

\begin{figure}[ht]
    \centering
\begin{subfigure}{0.49\textwidth}
    \centering
    \vtop{\null \hbox{\includegraphics[width=\textwidth]{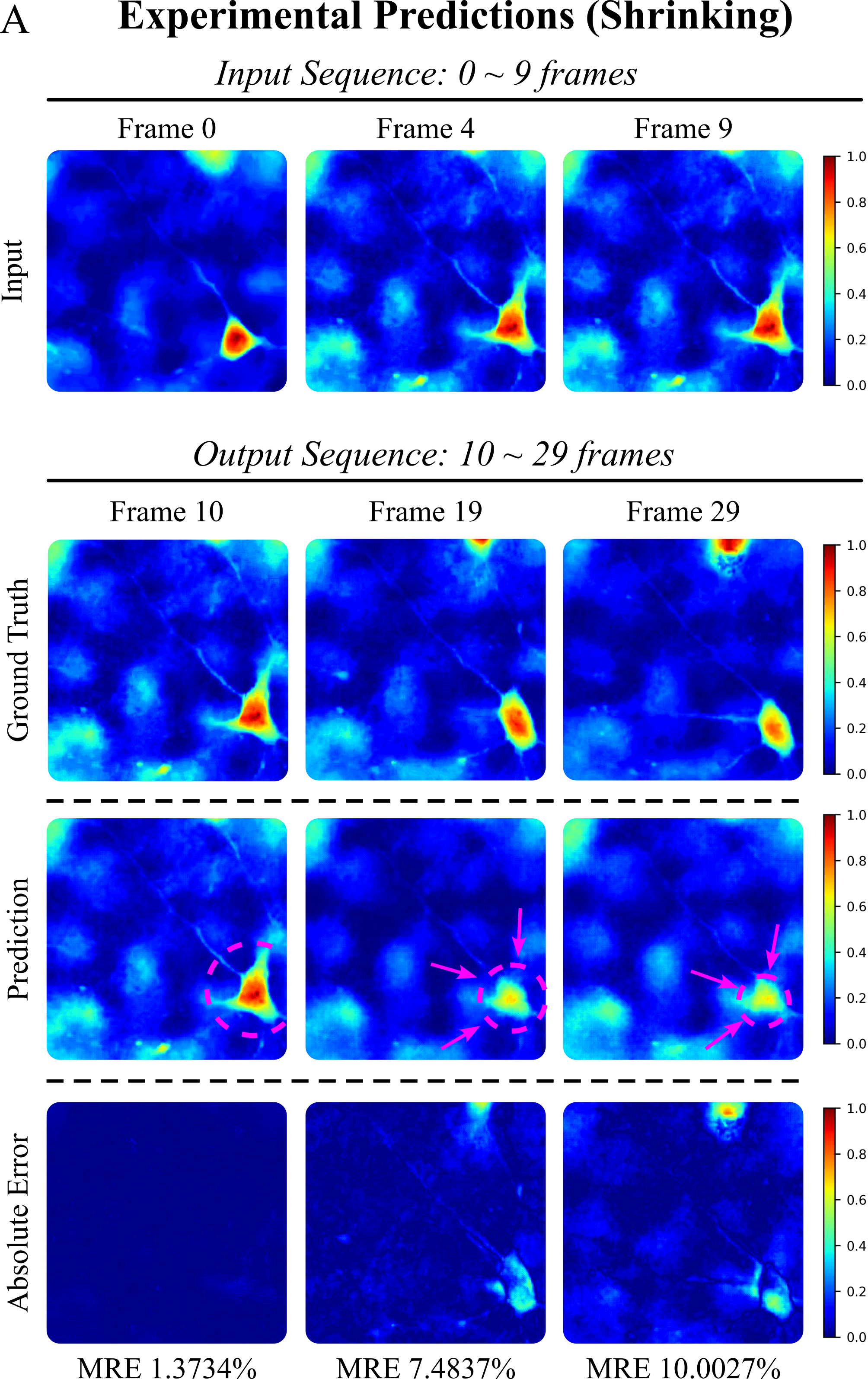}}}
\end{subfigure}
\hfill
\begin{subfigure}{0.49\textwidth}
    \centering
    \vtop{\null \hbox{\includegraphics[width=\textwidth]{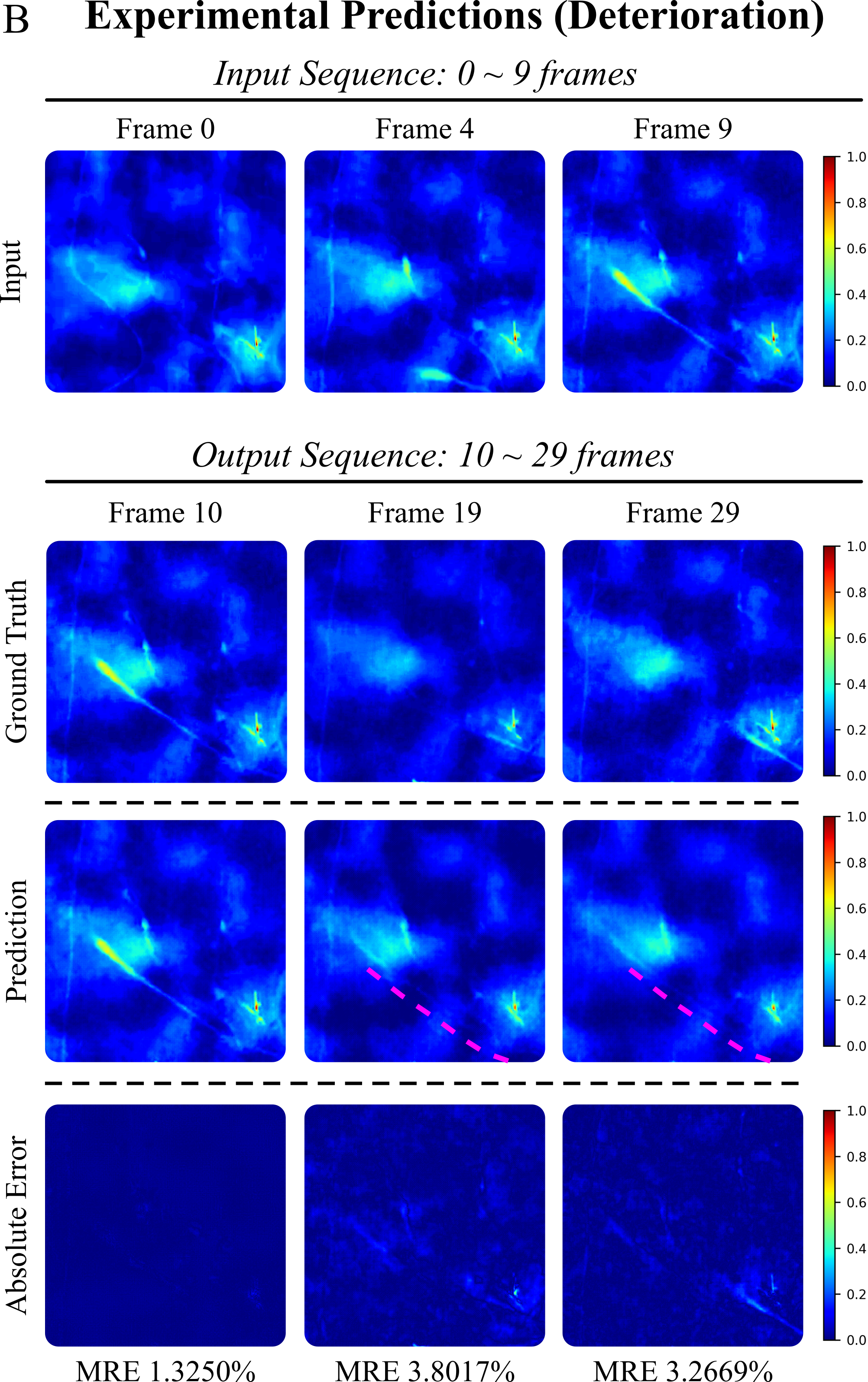}}}
\end{subfigure}
    \caption{(A\&B) Prediction for neurite shrinking and deterioration given experimental input. 
    The layout follows the same structure as Figure~\ref{fig:synthetic_predictions_1}.
    Magenta dashed circle, arrow, and line: the model is able to predict neurite deterioration, including shrinking soma and neurite deterioration exhibited in the experimental ground truth.}
    \label{fig:exp_connect_predictions}
\end{figure}

Figure~\ref{fig:exp_connect_predictions} presents model predictions for experimental neurite shrinking and deteriorating, following the same layout as previous figures. 
The results highlight the capability of our model to address complex morphological changes, including shrinking soma and neurite degeneration, with errors quantified by MRE values reaching up to 10.0027\% in later predictions. 
Magenta outlines, including dashed lines, circles, and arrows, annotate specific regions where the model effectively captures key transformations. 
Magenta circles identify areas in the predictions and ground truth where soma shrinkage and arrows indicate the direction of boundary evolution. 
The magenta dashed line highlights regions where neurite previously resided but disappeared during the experimental culturing process, indicating potential deterioration. 
Despite the experimental dataset introducing significant challenges, such as concurrent translation, scaling, deformation variations, and gradient-like noises possibly from the imaging process, the model consistently predicts neurite deterioration, demonstrating its ability to handle complex and dynamic changes.
These predictions made by the model based on real experimental data provide valuable insights that can help guide experimental culturing procedures in real-time. 
This offers researchers a fast computational tool to analyze morphological changes in advance, determine the success rate of the experiment, or help study neurite deterioration. 
In this manner, the digital twin framework has the potential to enhance experimental culturing procedures for neurons, optimize resource allocation, and aid the development of targeted treatment for NDDs.

\subsection{Prediction error and limitations}

\begin{figure}[ht]
\centering
\includegraphics[width=\textwidth]{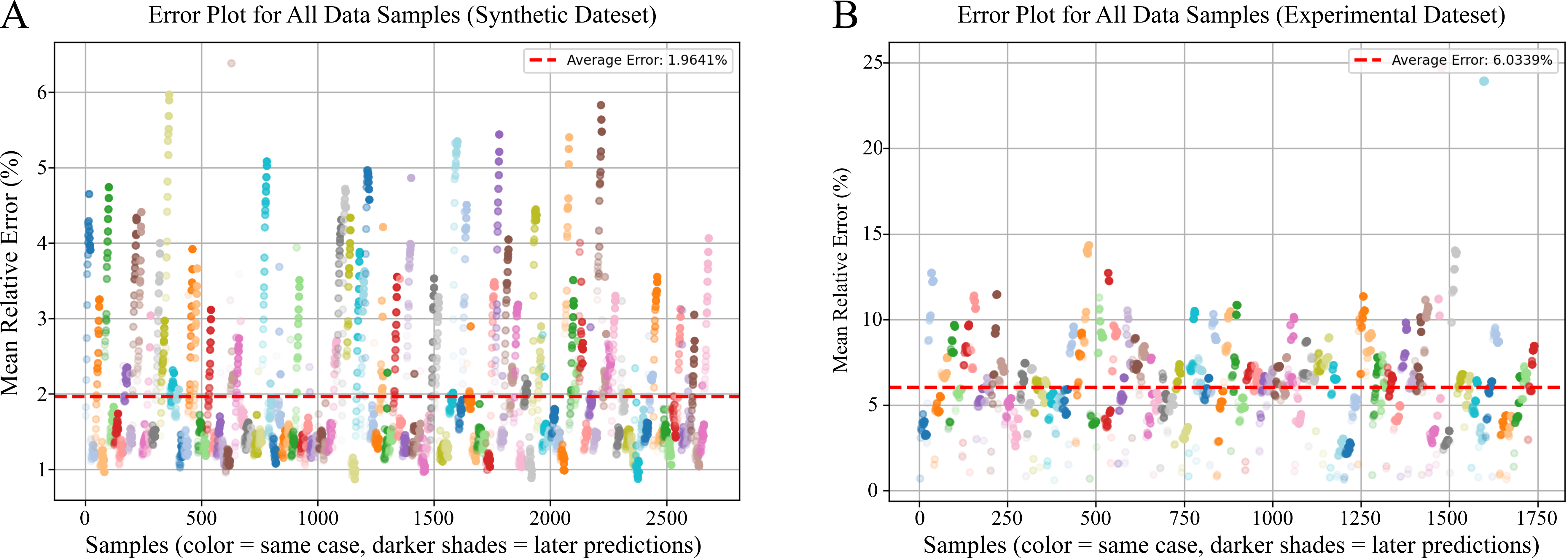}
    \caption{Comparison of error plots for (A) synthetic and (B) experimental datasets predictions.}
    \label{fig:test_error_plots}
\end{figure}

The prediction errors are summarized in Figure~\ref{fig:test_error_plots}, which showcases the MRE for each sample in the dataset. 
Each color in the scatter plot corresponds to a specific case, with darker shades representing predictions further into the future.
As predictions extend further into the future, a clear increasing trend in MRE is observed, highlighting the inherent challenges of long-term temporal forecasting.
The maximum error observed in Figure~\ref{fig:test_error_plots} is approximately 6.25\% for the synthetic dataset and 24\% for the experimental dataset. 
These errors occur because the experimental dataset is derived from microscopy images, which are inherently susceptible to imaging quality variations, such as focus and lighting conditions. 
Experimental neuron cultures are also 3D, whereas the model and imaging operate within 2D space. 
This dimensional mismatch may result in neurons unexpectedly disappearing from the image (out of focus or occluded by other objects), creating challenging prediction patterns. 
Furthermore, preprocessing steps applied to the experimental dataset (Section \ref{sec:datagen}), such as scaling and handling outliers, might inadvertently modify or exclude critical details, contributing to the observed errors. 
These factors contribute to the inherent challenges in working with real-world experimental data.
Despite this, the model achieves an overall average test error of 1.9641\% for synthetic neurite deterioration prediction and 6.0339\% for experimental neurite deterioration prediction, demonstrating that it effectively captures neurite deteriorations over extended time frames across both synthetic and experimental datasets. 
However, certain discrepancies, such as those observed in the absolute error maps, may be attributed to unaccounted external factors, variability in neurite morphology, or unexpected neuron interactions, such as the sudden entrance of another neuron highlighted by the magenta dashed circles in Figure~\ref{fig:experimental_neuron_predictions}.
These errors suggest potential ways to refine the model, such as incorporating additional parameters or leveraging more attention mechanisms to enhance spatiotemporal learning.

Ideally, the framework would work through a transfer learning~\cite{weiss2016survey} or a few-shot learning approach~\cite{wang2020generalizing}. 
Initially, the ML model would be pre-trained on the synthetic dataset to capture diverse neurite deterioration behaviors to overcome the limited experimental dataset size. 
The pre-trained model would then be fine-tuned on the limited experimental dataset, with selected layers frozen, enabling adaptations to real experimental scenarios while leveraging learned features. 
By integrating synthetic and experimental datasets, the two would complement each other, resulting in a more robust and capable ML-based digital twin framework.
However, the current framework handles synthetic and experimental datasets separately. 
This separation is due to significant differences in neurite behaviors between the datasets.
Our current experimental culture images exhibit complex morphological transformations, including translation, scaling, and rotation along with different artifacts from the imaging process, whereas the synthetic dataset only captures stationary neurons with neurite deterioration. 
These disparities interfere with model training when combined, leading to suboptimal performance.
With further work and effort, both the IGA-based phase field model and the experimental imaging process can be refined to generate datasets with comparable neurite morphometrics, enabling the full potential of this high-throughput digital twin framework.
From experimental culturing perspectives, this digital twin framework provides crucial insights that significantly enhance experimental design and throughput. 
By leveraging the digital twin framework, researchers can predict neurite culture behaviors and make informed decisions on the placement of neurons, extracellular conditions, or specific parameters for follow-up NDDs studies. 
This approach will reduce costly, time-intensive trial-and-error procedures and accelerate experimental validation for targeted treatment development.
Furthermore, the digital twin framework accelerates the validation of hypotheses on different neuron growth factors related to NDDs and the development of targeted treatments. 
The ability to simulate and predict complex neurite dynamics in real-time means that the digital twin can bridge the gap between computational models and experimental neuron culture to support faster and more accurate discoveries.

\section{Conclusion and future work}
\label{sec: conclusion}

In this paper, we introduce a high-throughput digital twin framework that integrates the IGA-based phase field model to generate synthetic training datasets and incorporates experimental data for enhanced model predictions. 
The framework is designed to predict neurite deterioration quickly and accurately, and it has extended applications in diverse experimental neuron cultures. 
It directly tackles the diverse neurite morphological characteristics and the scarcity of experimental data that are often costly and time-intensive to obtain.
By combining synthetic and experimental datasets, this framework enables efficient simulation and analysis of complex neurite dynamics while providing valuable insights to guide experimental culturing decisions. 
A novel synthetic dataset capturing biomimetic neurite deterioration patterns was developed to mitigate limitations in experimental data. 
The model utilizes a MetaFormer-based gSTA architecture to effectively capture intricate temporal dependencies and complex morphological transformations, ensuring accurate predictions.
With a combined VGG16-based perceptual and MSE loss, our approach achieves high prediction accuracy, with mean errors of 1.9641\% for synthetic datasets and 6.0339\% for experimental datasets.
This demonstrates its potential for advancing neurological disorder study and development for target treatment planning.

The current digital twin framework processes synthetic and experimental datasets separately, limiting its ability to fully leverage the complementary strengths of both.
The limited diversity of experimental datasets constrains the generalization of the model across varied scenarios. 
Addressing this limitation requires expanding dataset diversity and adopting fine-tuning strategies to improve robustness and adaptability.
The ML architecture currently employs a VGG-based perceptual loss pre-trained on ImageNet, which assumes feature alignment with neuron image data but may not fully reflect the actual neurite features in the neuron dataset. 
Furthermore, experimental datasets often exhibit complex dynamics, including drastic and unpredictable movements and deformations, which are absent in synthetic data and present challenges for the current digital twin framework. 

Future work will focus on aligning the characteristics of synthetic and experimental datasets to bridge the gap between IGA-based phase field model simulations and real-world experimental neuron cultures. 
Developing domain-specific perceptual loss models tailored to neuronal images will enable the model to capture intricate details and nonlinear behaviors in the dataset.
This will enhance the accuracy and relevance for complex experimental conditions.
Future extensions could involve adapting the model to simulate more complex neuron networks, encompassing intricate interactions between multiple neurons and diverse morphologies. 
Incorporating multimodal data could also further expand the predictive capabilities and broaden its applicability. 
In addition, exploring the usage in other biological processes, such as axonal regeneration or synapse formation, may open up new avenues in neuroscience. 
By improving temporal prediction accuracy and capturing morphological changes, this approach can advance neuroscience research while supporting clinical neurobiology applications, including early diagnostics, target treatment planning, and therapeutic interventions.

\section*{Code and data availability}
The code and datasets generated and analyzed in this paper are accessible in the ``NDD\_ML" GitHub repository. \url{https://github.com/CMU-CBML/NDD_ML} (DOI:\href{https://doi.org/10.5281/zenodo.14502515}{10.5281/zenodo.14502516}).
Correspondence and requests for code and data should be addressed to K. Qian or Y. J. Zhang.

\section*{Declaration of competing interest}
The authors declare no known competing financial interests or personal relationships that could have appeared to influence the work reported in this paper.

\section*{Acknowledgement}
K. Qian and Y. J. Zhang were supported in part by the NSF grants CMMI-1953323 and CBET-2332084.
This work used RM-nodes on Bridges-2 Supercomputer at Pittsburgh Supercomputer Center \cite{ecss,xsede} through allocation ID eng170006p from the Advanced Cyberinfrastructure Coordination Ecosystem: Services \& Support (ACCESS) program, which is supported by National Science Foundation grants \#2138259, \#2138286, \#2138307, \#2137603, and \#2138296.

\bibliographystyle{elsarticle-num}
\bibliography{reference}
\end{document}